\documentclass[apj]{emulateapj}

\newcommand{\ms}{\mbox{m\,s$^{-1}$}}
\newcommand{\kms}{\mbox{km \ s$^{-1}$}}
\newcommand{\Msun}{\mbox{M$_{\odot}$}}
\newcommand{\Rsun}{\mbox{R$_{\odot}$}}
\newcommand{\Mjup}{\mbox{M$_{\rm J}$}}
\newcommand{\angst}{\mbox{${\rm \AA}$}}

\slugcomment{}

\shorttitle{A Pair of Giant Planets around HD 47366}
\shortauthors{Sato et al.}

\begin{document}


\title{A Pair of Giant Planets around the Evolved Intermediate-Mass Star HD 47366:
Multiple Circular Orbits or a Mutually Retrograde Configuration}


\author{Bun'ei Sato\altaffilmark{1},
Liang Wang\altaffilmark{2},
Yu-Juan Liu\altaffilmark{2},
Gang Zhao\altaffilmark{2},
Masashi Omiya\altaffilmark{3},
Hiroki Harakawa\altaffilmark{3},
Makiko Nagasawa\altaffilmark{4},
Robert A.~Wittenmyer\altaffilmark{5,6},
Paul Butler\altaffilmark{7},
Nan Song\altaffilmark{2},
Wei He\altaffilmark{2},
Fei Zhao\altaffilmark{2},
Eiji Kambe\altaffilmark{8},
Kunio Noguchi\altaffilmark{3},
Hiroyasu Ando\altaffilmark{3},
Hideyuki Izumiura\altaffilmark{8,9},
Norio Okada\altaffilmark{3},
Michitoshi Yoshida\altaffilmark{10},
Yoichi Takeda\altaffilmark{3,9},
Yoichi Itoh\altaffilmark{11},
Eiichiro Kokubo\altaffilmark{3,9},
and Shigeru Ida\altaffilmark{12}
}
\email{satobn@geo.titech.ac.jp}

\altaffiltext{1}{Department of Earth and Planetary Sciences, Tokyo Institute of
Technology, 2-12-1 Ookayama, Meguro-ku, Tokyo 152-8551, Japan}
\altaffiltext{2}{Key Laboratory of Optical Astronomy, National Astronomical Observatories,
   Chinese Academy of Sciences, Beijing 100012, China}
\altaffiltext{3}{National Astronomical Observatory of Japan, 2-21-1 Osawa,
   Mitaka, Tokyo 181-8588, Japan}
\altaffiltext{4}{Department of Physics, Kurume University School of Medicine,
67 Asahi-machi, Kurume-city, Fukuoka 830-0011, Japan}
\altaffiltext{5}{School of Physics, University of New South Wales, 
Sydney 2052, Australia}
\altaffiltext{6}{Australian Centre for Astrobiology, University of New 
South Wales, Sydney 2052, Australia}
\altaffiltext{7}{Department of Terrestrial Magnetism, Carnegie 
Institution of Washington, 5241 Broad Branch Road, NW, Washington, DC 
20015-1305, USA}
\altaffiltext{8}{Okayama Astrophysical Observatory, National
  Astronomical Observatory of Japan, Kamogata,
  Okayama 719-0232, Japan}
\altaffiltext{9}{The Graduate University for Advanced Studies,
  Shonan Village, Hayama, Kanagawa 240-0193, Japan}
\altaffiltext{10}{Hiroshima Astrophysical Science Center, Hiroshima University,
  Higashi-Hiroshima, Hiroshima 739-8526, Japan}
\altaffiltext{11}{Nishi-Harima Astronomical Observatory, Center for Astronomy,
University of Hyogo, 407-2, Nishigaichi, Sayo, Hyogo
679-5313, Japan}
\altaffiltext{12}{Earth-Life Science Institute, Tokyo Institute of
Technology, 2-12-1 Ookayama, Meguro-ku, Tokyo 152-8551, Japan}




\begin{abstract}
We report the detection of a double planetary system around
the evolved intermediate-mass star HD 47366 from precise radial-velocity
measurements at Okayama Astrophysical Observatory, Xinglong
Station, and Australian Astronomical Observatory.
The star is a K1 giant with a mass of 1.81$\pm0.13~M_{\odot}$, a radius
of $7.30\pm0.33~R_{\odot}$, and solar metallicity.
The planetary system is composed of two giant planets
with minimum mass of $1.75^{+0.20}_{-0.17}~\Mjup$ and
$1.86^{+0.16}_{-0.15}~\Mjup$,
orbital period of $363.3^{+2.5}_{-2.4}$ d and $684.7^{+5.0}_{-4.9}$ d,
and eccentricity of $0.089^{+0.079}_{-0.060}$ and $0.278^{+0.067}_{-0.094}$,
respectively, which are derived by a double Keplerian orbital fit to the
radial-velocity data. The system adds to the population of multi-giant-planet
systems with relatively small orbital separations, which are
preferentially found around evolved intermediate-mass stars.
Dynamical stability analysis for the system revealed, however, that the best-fit orbits are
unstable in the case of a prograde configuration.
The system could be stable if the planets were in 2:1 mean-motion resonance,
but this is less likely considering the observed period ratio and eccentricity.
A present possible scenario for the system is that both of the planets have nearly circular
orbits, namely the eccentricity of the outer planet is less than $\sim$0.15, which is just within
$1.4~\sigma$ of the best-fit value, or the planets are in a mutually retrograde
configuration with a mutual orbital inclination larger than 160$^{\circ}$.
\end{abstract}





\section{Introduction}\label{intro}

Evolved stars (giants and subgiants) have been extensively
searched for the last ten years aiming to explore planets around more massive stars than
the Sun and to investigate orbital evolution of planetary systems during the post
main-sequence phase. Precise radial-velocity surveys have discovered about 120
substellar companions around such evolved stars\footnote{Evolved stars here include
those with logarithmic surface gravity $\log{g}<4.0$. The list of the stars harboring substellar
companions is from NASA Exoplanet Archive.} so far, and they are now known to exhibit
statistical properties that are not necessarily similar to those around solar-type stars
\citep[e.g.,][]{setiawan:2005, hatzes:2005, hatzes:2006, lovis:2007,
dollinger:2009, demedeiros:2009, johnson:2011, wittenmyer:2011,
wang:2012, omiya:2012, sato:2013b, novak:2013, lee:2014, trifonov:2014, wang:2014,
jones:2014, reffert:2015}.
Recently, very high-precision photometry by {\it Kepler} space telescope has
succeeded in detecting planetary transits on evolved stars \citep{batalha:2013}.
The discoveries include short-period planets and sub-Jupiter-mass ones
around giants \citep[][]{huber:2013, lillobox:2014, quinn:2015, ciceri:2015, ortiz:2015, sato:2015},
which had rarely been found by the
ground-based radial-velocity surveys, and also candidates with longer periods,
which could be counterparts of those found by the radial-velocity surveys
\footnote{NASA Exoplanet Archive}.

\begin{figure}
\epsscale{1.2}
\plotone{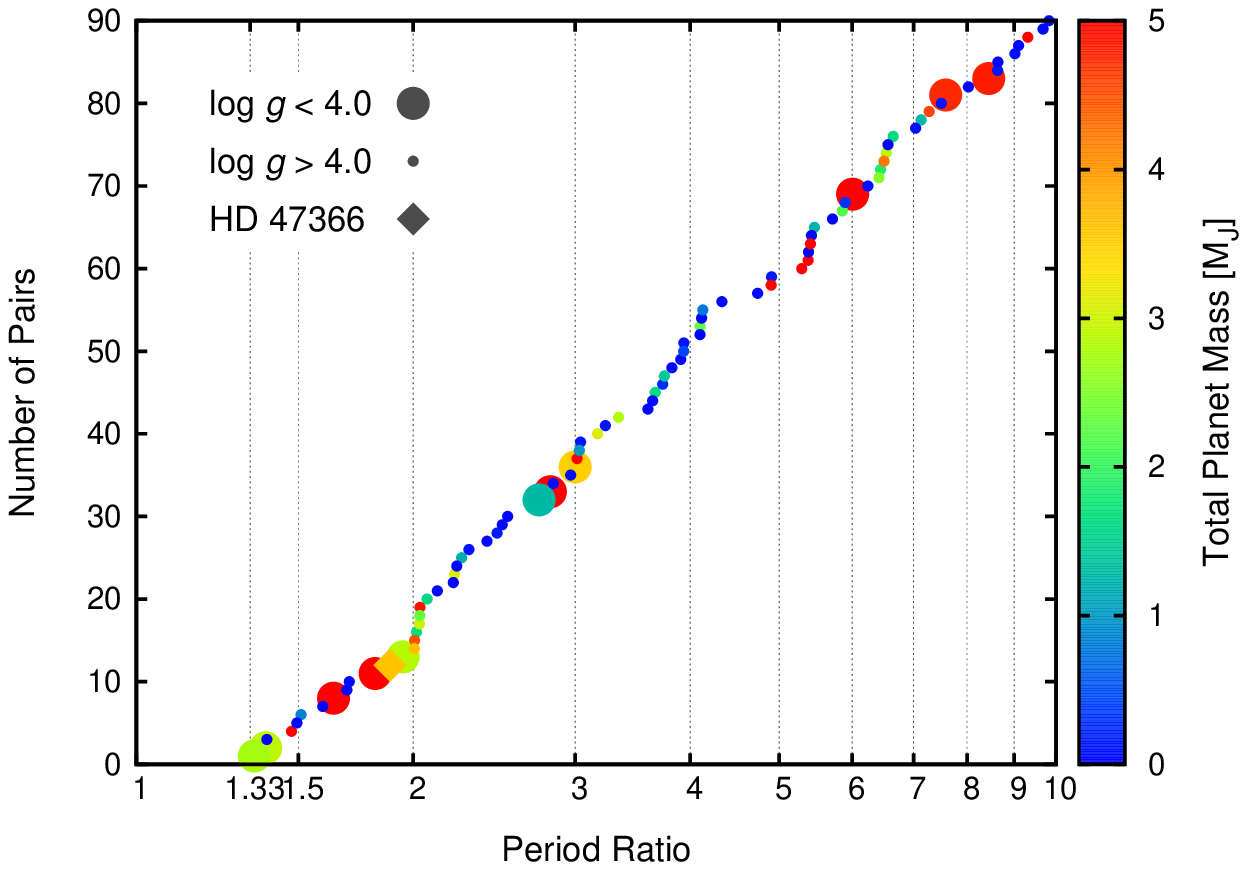}
\plotone{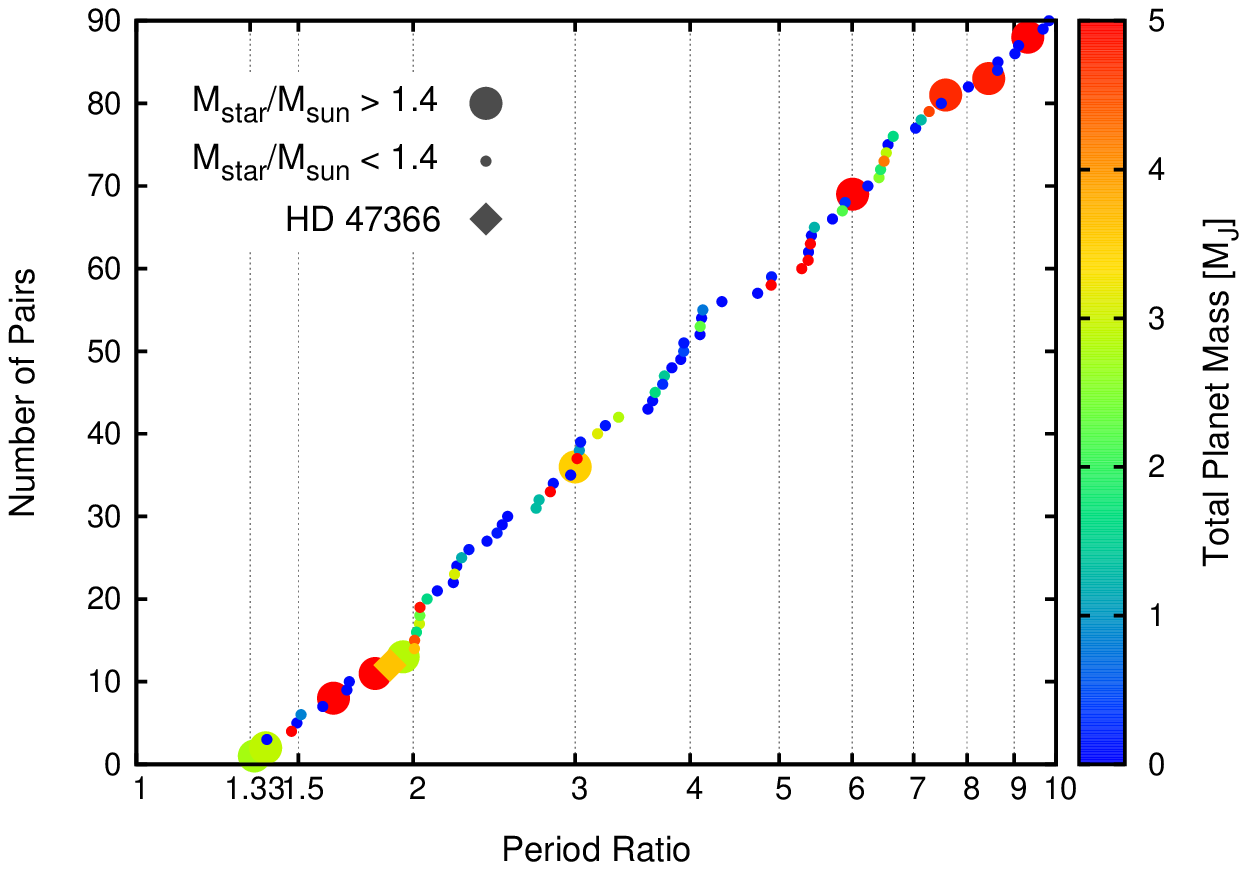}
\caption{Cumulative number of planet pairs in multiple-planetary systems discovered with
ground-based radial-velocity and transit surveys.  The pairs with period ratio smaller than
10 are plotted.
The color indicates the total mass of the planet pair and the symbol size indicates the
$\log~g$ ({\it top}) and mass ({\it bottom}) of the host star. The data were downloaded from the
NASA Exoplanet Archive. Since the ongoing Doppler planet searches target at once evolved
and intermediate-mass stars, the two panels look nearly identical.}
\label{period-ratio}
\end{figure}

Among the planetary systems around evolved stars, multiple-planet ones
are of particular interest in terms of formation and evolution of planetary systems.
Figure \ref{period-ratio} shows the cumulative number of planet pairs in multiple-planet
systems discovered by ground-based radial-velocity and transit surveys.
The total mass of the planet pair, and the $\log~g$ and mass of the host star are
indicated in color and size of the symbol, respectively. Interestingly almost all of the
giant-planet pairs with period ratio smaller than 2 are preferentially found around
evolved intermediate-mass stars.\footnote{One exception is the planetary system
around the K2.5 V star HD 45364, which consists of two giant planets
($m_b\sin i=0.187~M_{\rm J}$ and $m_c\sin i=0.658~\Mjup$)  in
3:2 resonance \citep{correia:2009}.}
For example, 24 Sex \citep[G5\,IV;][]{johnson:2011} and $\eta$ Cet \citep[K2\,III;][]{trifonov:2014}
host double giant-planet systems probably in mean-motion resonance of 2:1, which
suggests that these systems have experienced differential convergent orbital migration
\citep[e.g.,][]{lee:2002, kley:2004}.
In contrast, double giant-planet systems around HD 200964 \citep[K0\,IV;][]{johnson:2011}
and HD 5319 \citep[G5\,IV;][]{robinson:2007,giguere:2015} are near the 4:3 resonance,
which are considered to be difficult to form either via convergent migration,
scattering or in situ formation \citep{rein:2012}.
BD+20 2457 (K2\,III) hosts two brown dwarfs that are suggested to be near
3:2 resonance \citep{niedzielski:2009}. However, detailed orbital stability analysis
for the system revealed that the best-fit orbits derived by radial-velocity data are unstable,
while it would be stable if the two brown dwarfs are mutually in retrograde
configuration \citep{horner:2014}. Giant-planet pairs with such small period ratio
are rarely found around solar-type stars, though many less massive ones
including Neptunes and Super-Earths are found around them.

Here we report the detection of a double giant-planet system around the intermediate-mass
giant HD 47366 (K1 III, $M=1.8~\Msun$),
which adds to the growing population of multi-planet systems with small period ratio
around evolved stars. The discovery was made by Okayama and Xinglong Planet Search Programs
\citep[e.g.,][]{wang:2012, sato:2013b} facilitated by joint precise radial-velocity
observations from Australian Astronomical Observatory \citep{sato:2013a}.
The planetary system is intriguing in the points that the best-fit Keplerian orbit is unstable,
it is near but less likely in 2:1 mean-motion resonance, and could be stable
if the orbits are nearly circular or in retrograde configuration.

The rest of the paper is organized as follows. The stellar properties are presented in section
\ref{stpara} and the observations are described in section \ref{obs}. Orbital solution and
results of dynamical stability analysis are presented in section \ref{ana} and \ref{dynamics},
respectively.
Section \ref{summary} is devoted to discussion and summary.

\begin{figure}
    \plotone{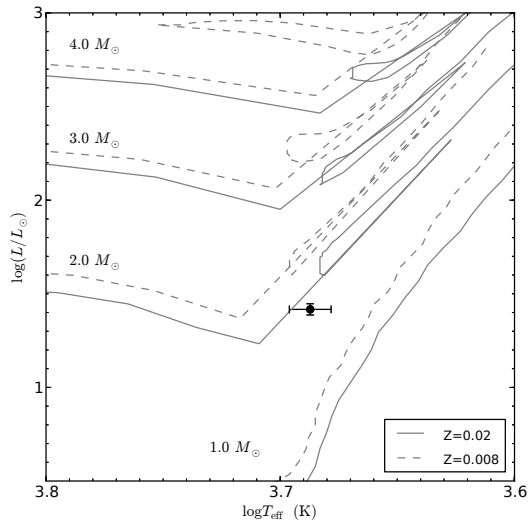}\epsscale{1.2}
    \caption{
        HD\,47366 (solid circle) on the H-R diagram.
        The error bars correspond to the derived uncertainties of $\log{L}$ and
            $T_{\rm eff}$.
        The solid and dashed lines reprensent the evolutionary tracks from
            \citet{lejeune:2001} for stars of $M=1\sim4\,M_\odot$ with $Z=0.02$
            (solar metallicity) and $Z=0.008$, respectively.
    }\label{fig-hrd}
\end{figure}


\begin{figure}
\plotone{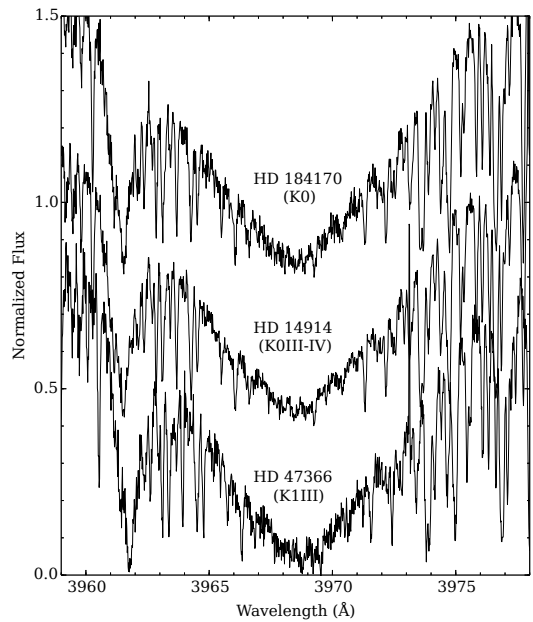}\epsscale{1.2}
\caption{Spectra in the region of Ca II H lines. Stars with similar
spectral type to HD 47366 in our sample are also shown.
A vertical offset of about 0.4 is added to each spectrum.}
\label{fig:CaIIH}
\end{figure}

\begin{deluxetable*}{lccc}
\tablecaption{Stellar Parameters for HD 47366\label{tbl:star}}
\tablewidth{0pt}
\tablehead{
\colhead{Parameter} & \colhead{This work} & \colhead{\citealt{Liu2010}} & \colhead{\citealt{Mishenina2006}}}
\startdata
        Spectral Type                    & K1\,III          &     & \\
        Hipparcos Parallax $\pi$ (mas)                 & $12.5\pm0.42$    &     & \\
        Distance (pc)               & $80.0\pm2.7$     &     & \\
        Visual Magnitude $V$                         & 6.11             &     & \\
        Color Index $B-V$                       & $0.994\pm0.002$  &     & \\
        Redenning $E(B-V)$                    & 0.028            &     & \\
        Interstellar Extinction $A_V$                       & 0.087            &     & \\
        Absolute Magnitude $M_V$                       & 1.51             & 1.467          & 1.044 \\
        Bolometric Correction $B.C.$                        & $-0.309$         & $-0.302$       &       \\
        Bolometric Magnitude $M_{\rm bol}$               & 1.20             & 1.165          &       \\
        Effective Temperature $T_{\rm eff}$ (K)           & $4866\pm100$     & $4883\pm100$   & 4772  \\
        Surface Gravity $\log g$ (cgs)                    & $2.97\pm0.06$    & $3.00\pm0.1$   & 2.60  \\                                                      
        Metallicity [Fe/H]                      & $-0.02\pm0.09$   & $-0.10\pm0.1$  & $-0.16$ \\
        Microturbulent Velocity $v_{\rm t}$ (km\,s$^{-1}$)  & 1.24             & $1.2\pm0.2$    & 1.2   \\
        Macroturbulent Velocity $v_{\rm macro}$ (km\,s$^{-1}$) & $5.54\pm0.45$    &                &       \\
        Projected Rotational Velocity $v\sin{i}$ (km\,s$^{-1}$)   & $4.3\pm0.8$      &                &       \\
        Luminosity $L\ (L_\odot)$              & $26.1\pm1.8$     & $26.2$         & 38.5  \\
        Radius $R\ (R_\odot)$              & $7.30\pm0.33$    & 7.16           & 9.09  \\
        Mass $M\ (M_\odot)$              & $1.81\pm0.13$    & 1.87           & 1.2   \\
        Age (Gyr)                   & $1.61\pm0.53$    &                &       \\
\enddata
\end{deluxetable*}

\section{Stellar Properties}\label{stpara}

HD\,47366 (HIP\,31674, BD\,$-12$\,1566, HR\,2437, TYC\,5373-2001-1) is listed in
    the {\sc Hipparcos Catalogue} \citep{ESA:1997} as a K1\,III star,
    with a visual magnitude of $V=6.11$ and a color index of $B-V=0.994$.
The Hipparcos parallax $\pi=12.5\pm0.42$\,mas \citep{vanLeeuwen2007} corresponds
    to a distance of $80.0\pm2.7$ pc.
The reddening $E(B-V)=0.028$ was obtained from the Galactic dust map of
    \citet{Schlegel1998}, with the correction given by \citet{Bonifacio2000} and
    a scaling factor of $1-\exp{(-|d\sin{b}|/h)}$, where $d$ is the distance,
    $b$ is the Galactic latitude and $h=125$ pc is the scale height of the
    reddening layer.
The absolute magnitude $M_V=1.51$ was derived from the distance and the
    interstellar extinction $A_V=3.1E(B-V)$.
By adopting the broad-band photometric color $B-V$ and the estimated metallicity
    with the empirical calibration relation of \citet{Alonso1999, Alonso2001},
    we derived the bolometric correction $BC=-0.309$ and the effective
    temperature $T_{\rm eff}=4866\pm100$ K.
We used a high signal-to-noise ratio (S/N$\sim$200), iodine-free spectrum taken
    with HRS to measure the equivalent widths (EWs) of $\sim$30 Fe I lines, to
    derive the iron abundance [Fe/H].
The line list as well as their oscillation strengths ($\log{gf}$) were mainly
    taken from \cite{hekker:2007}, in which iron lines were carefully selected to
    avoid any blend by atomic or CN lines.
The model atmosphere used in this work was interpolated from the line-blanketed,
    local thermodynamic equilibrium (LTE) ATLAS9-ODFNEW grid
    \citep{Castelli2004}.
The microturbulent velocity $v_{\rm t}$ was obtained by minimizing the trend between the
    abundances of different Fe I lines and their reduced equivalent widths
    ($\log({\rm EWs}/\lambda)$).
The macroturbulent velocity was estimated with the empirical relations of
    $v_{\rm macro}$ v.s. $T_{\rm eff}$ given by \citet{hekker:2007}, and the
    projected rotational velocity ($v\sin{i}$) was determined with method of
    \citet{Fekel1997}.
The stellar mass, surface gravity ($\log{g}$), radius and age were derived
    using a Bayesian approach with the Geneva database \citep{lejeune:2001},
    which includes the post-helium flash phases for stars with
    $M\ge1.7\,M_\odot$.
Firstly, we interpolated an extensive grid of evolutionary tracks, with
    $\Delta M=0.05$ within $1.2\le M/M_\odot\le3.6$, $\Delta$[Fe/H] = 0.02
    within $-0.4\le$ [Fe/H] $\le+0.3$, and 500 points in each track.
Then for each data point, we calculated the likelihood functions of $\log{L}$,
    $T_{\rm eff}$ and [Fe/H] to match the observed values by assuming Gaussian
    errors.
We adopted uniform prior probabilities of mass and [Fe/H].
It is noted that stars evolving more slowly have higher probability of being observed.
Without correcting this evolution effect, the resulting parameters would bias towards
 the rapid evolution phases.
We therefore weighted the probability of each point along its evolutionary track by
the the normalized time-step $(a_{i+1,j} - a_{i,j})/(a_{n,j}-a_{a,j})$, where $a_{i,j}$
is the age of the $i$-th interpolated point in the $j$-th track, and $n=500$ is the number
of interpolated points in each track.
Eventually, the probability distribution functions (PDFs) of the parameters
    yield $M=1.81\pm0.13\,M_\odot$, $R=7.30\pm0.33\,R_\odot$,
    $\log{g}=2.97\pm0.06$, and age $=1.61\pm0.53$ Gyr.
The stellar mass is particularly important to derive the minimum masses of the
    orbiting planets detected with Doppler technique.
However, previous spectroscopic analyses gave discrepant results (1.87 $M_\odot$
    by \citealt{Liu2010}; 1.2 $M_\odot$ by \citealt{Mishenina2006}) for
    HD\,47366, which may due to the different methods on finding $T_{\rm eff}$
    and $\log{g}$.
Our determinations were based on a similar method to that of \cite{Liu2010}, but
    used the Geneva evolutionary tracks, instead of Y$^2$ model \citep{Yi2003}
    which does not take account of the evolutionary phases after the
    helium-core flash.
We found the probability that the star has passed through the RGB tip and in
    core helium burning phase is $\sim$88\%.
The stellar parameters of HD\,47366 are listed in table \ref{tbl:star}.
In figure \ref{fig-hrd}, we plotted HD\,47366 on the H-R diagram, together with
    the evolutionary tracks for stars with different masses and metal contents.

The star shows no significant emission in the core of Ca II HK lines as shown
in figure \ref{fig:CaIIH},
which suggests that the star is chromospherically inactive. We did not analyze
flux variations of Ca II HK lines for the star because of the poor S/N
ratio for the core of the lines.

\section{Observations}\label{obs}

\subsection{OAO Observations}

Observations of HD 47366 at OAO were made with the 1.88-m reflector and
the HIgh Dispersion Echelle Spectrograph \citep[HIDES;][]{izumiura:1999}
from December 2006 to April 2014. We used both of the conventional slit
mode (HIDES-S) and the high-efficiency fiber-link mode (HIDES-F) of the
spectrograph that became available since 2010 \citep{kambe:2013}.
In the case of the HIDES-S,  a slit width of the spectrograph was set to
200 $\mu$m ($0^{\prime\prime}.76$) corresponding to a spectral resolution
($R=\lambda/\Delta\lambda$) of 67000 by about 3.3 pixels sampling.
In the case of HIDES-F, a width of the image sliced by an image slicer
is $1^{\prime\prime}.05$, corresponding to a spectral resolution of
$R=55000$ by 3.8-pixel sampling. Each observing mode uses its own
iodine absorption cell for precise radial-velocity measurements
\citep[I$_2$ cell;][]{kambe:2002,kambe:2013}, which provides a fiducial
wavelength reference in a wavelength range of 5000--5800${\rm \AA}$.
We have obtained a total of 50 and 7 data points of HD 47366 using HIDES-S
and HIDES-F mode, respectively.
The reduction of echelle data (i.e. bias subtraction, flat-fielding,
scattered-light subtraction, and spectrum extraction) was performed
using the IRAF\footnote{IRAF is distributed by the National
Optical Astronomy Observatories, which is operated by the
Association of Universities for Research in Astronomy, Inc. under
cooperative agreement with the National Science Foundation,
USA.} software package in the standard way.

We performed radial-velocity analysis following the method of \citet{sato:2002},
\citet{sato:2012} and \citet{butler:1996}, in which an I$_2$-superposed spectrum
is modeled as a product of a high resolution I$_2$ and a stellar template spectrum
convolved with a modeled point spread function (PSF) of the spectrograph.
We model the PSF using multiple gaussian profiles \citep{val:95} and obtain
the stellar spectrum by deconvolving a pure stellar spectrum with the spectrograph
PSF estimated from an I$_2$-superposed B-star spectrum.
We obtained a typical measurement error in radial-velocity of about 4--5 m s$^{-1}$
for the star, which was estimated from an ensemble of velocities from each of
$\sim$300 spectral chunks (each $\sim$3${\rm \AA}$ long) in every exposure.
We listed the derived radial velocities for OAO data
in Table~\ref{OAOvels} together with the estimated uncertainties.

\LongTables
\begin{deluxetable}{lccc}
\tablecolumns{4}
\tablewidth{0pt}
\tablecaption{OAO Radial Velocities for HD 47366}
\tablehead{
\colhead{JD$-$2450000} & \colhead{Velocity (\ms)} & \colhead{Uncertainty (\ms)} & \colhead{Mode}}
\startdata
\label{OAOvels}
4093.18674 & 11.36 & 4.59 & HIDES-S \\ 
4144.11514 & $-$19.66 & 4.69 & HIDES-S \\ 
4172.01541 & $-$18.29 & 3.85 & HIDES-S \\ 
4467.12999 & $-$4.46 & 4.31 & HIDES-S \\ 
4553.97104 & $-$31.25 & 4.14 & HIDES-S \\ 
4755.29094 & 32.92 & 3.46 & HIDES-S \\ 
4796.25711 & 36.33 & 4.17 & HIDES-S \\ 
4818.17076 & 19.49 & 3.87 & HIDES-S \\ 
4823.11554 & 14.96 & 4.17 & HIDES-S \\ 
4834.16358 & 12.38 & 4.36 & HIDES-S \\ 
4856.08725 & $-$6.42 & 3.92 & HIDES-S \\ 
4863.09739 & $-$6.67 & 4.41 & HIDES-S \\ 
4864.03798 & $-$11.43 & 3.60 & HIDES-S \\ 
4883.04016 & $-$7.18 & 4.55 & HIDES-S \\ 
4923.93523 & $-$17.64 & 3.73 & HIDES-S \\ 
5108.31507 & 40.77 & 4.18 & HIDES-S \\ 
5131.32409 & 47.90 & 3.57 & HIDES-S \\ 
5161.34584 & 39.99 & 4.64 & HIDES-S \\ 
5182.16055 & 11.19 & 4.10 & HIDES-S \\ 
5204.11564 & $-$7.29 & 4.29 & HIDES-S \\ 
5232.99361 & $-$43.84 & 3.62 & HIDES-S \\ 
5267.01163 & $-$60.09 & 3.57 & HIDES-S \\ 
5294.93562 & $-$62.39 & 3.54 & HIDES-S \\ 
5471.29879 & 24.76 & 3.61 & HIDES-S \\ 
5502.23739 & 33.89 & 4.15 & HIDES-S \\ 
5525.22640 & 33.98 & 4.85 & HIDES-S \\ 
5545.21850 & 18.86 & 3.73 & HIDES-S \\ 
5581.04001 & $-$12.92 & 3.75 & HIDES-S \\ 
5611.01311 & $-$8.59 & 4.13 & HIDES-S \\ 
5625.02350 & $-$13.62 & 3.49 & HIDES-S \\ 
5656.93188 & $-$10.50 & 3.74 & HIDES-S \\ 
5854.27845 & 33.84 & 3.69 & HIDES-S \\ 
5877.25348 & 18.63 & 5.42 & HIDES-S \\ 
5882.29779 & 12.23 & 3.83 & HIDES-S \\ 
5922.13357 & $-$12.73 & 3.88 & HIDES-S \\ 
5938.09758 & $-$34.44 & 3.79 & HIDES-S \\ 
5976.03887 & $-$49.42 & 4.57 & HIDES-S \\ 
6010.93116 & $-$64.77 & 5.74 & HIDES-S \\ 
6032.96687 & $-$55.43 & 4.48 & HIDES-S \\ 
6215.30762 & 27.47 & 4.04 & HIDES-S \\ 
6235.27265 & 41.83 & 3.68 & HIDES-S \\ 
6249.30138 & 27.36 & 4.69 & HIDES-S \\ 
6284.08997 & $-$0.00 & 4.69 & HIDES-S \\ 
6552.29701 & 42.87 & 5.53 & HIDES-S \\ 
6577.28414 & 4.35 & 4.84 & HIDES-S \\ 
6578.28739 & 14.21 & 4.48 & HIDES-S \\ 
6578.30265 & 14.55 & 4.27 & HIDES-S \\ 
6579.25810 & 12.00 & 5.63 & HIDES-S \\ 
6618.19944 & 4.11 & 3.91 & HIDES-S \\ 
6753.98409 & $-$63.39 & 4.45 & HIDES-S \\
\hline
6609.21551 & 21.82 & 4.04 & HIDES-F \\ 
6616.27916 & 33.33 & 3.44 & HIDES-F \\ 
6664.09442 & $-$1.41 & 4.02 & HIDES-F \\ 
6672.05938 & $-$16.16 & 3.85 & HIDES-F \\ 
6697.15561 & $-$9.17 & 4.70 & HIDES-F \\ 
6713.98988 & $-$17.94 & 4.87 & HIDES-F \\ 
6727.00570 & $-$12.14 & 3.47 & HIDES-F
\enddata
\end{deluxetable}
\subsection{Xinglong Observations}

Observations of HD 47366 at Xinglong station started in November 2006 using the
2.16-m reflector and the Coud${\acute{\rm e}}$ Echelle Spectrograph \citep[CES;][]{zhao:2001}.
We used the blue-arm, middle focal length camera, and 1K$\times$1K CCD (pixel size of
24$\times$24$\mu$m$^2$; hereafter CES-O) of the spectrograph, which covered a wavelength
range of 3900--7260$\angst$ with a spectral resolution of 40,000 by 2 pixel sampling.
Although a wide wavelength range was obtained with a single exposure, only a wave
band of $\Delta\lambda\sim$470$\angst$ was available for radial-velocity measurements
using an I$_2$ cell because of the small format of the CCD. 
Radial-velocity analysis for CES data was performed by optimized method of \citet{sato:2002}
for CES \citep{liu:2008}.
We took five data points for HD 47366 using CES-O from November 2006 to
December 2007 whose radial-velocity uncertainties were better than 25 m s$^{-1}$.
In March 2009, the old CCD was replaced by a new CCD having a smaller pixel size
(2K$\times$2K, 13$\times$13$\mu$m$^2$), and it gave a better radial-velocity precision
than before, although the wavelength coverage was unchanged \citep{wang:2012}.
We made observations of HD 47366 using the new CCD (hereafter CES-N) from December 2009
to December 2010, and collected a total of 26 data points with a radial-velocity precision
of 12--25 m s$^{-1}$ depending on weather condition.

Since November 2011, we have observed HD 47366 with the newly developed High Resolution
Spectrograph (HRS) attached at the Cassegrain focus of the 2.16m telescope.
The fiber-fed spectrograph is the successor of the CES, giving higher wavelength resolution
and throughput. The single 4K$\times$4K CCD provides a simultaneous wavelength coverage
of 3700--9200$\angst$, and the slit width of 190$\mu$m gives a wavelength resolution $R=45000$
with 3.2 pixels sampling. An iodine cell is installed before the fiber entrance at the Cassegrain focus.
We collected a total of 60 observations for the star from November 2011 to March 2014.
Radial-velocity analysis for the HRS data was performed by the same method as \citet{sato:2002}
and \citet{sato:2012}, but optimized for HRS. The resulting measurement errors in radial velocity
are ranging from 6 to 20 m s$^{-1}$ depending on weather condition.
The derived radial velocities for Xinglong data are listed
in Table~\ref{Xingvels} together with the estimated uncertainties.

\LongTables
\begin{deluxetable}{lccc}
\tabletypesize{\scriptsize}
\tablecolumns{4}
\tablewidth{0pt}
\tablecaption{Xinglong Radial Velocities for HD 47366}
\tablehead{
\colhead{JD$-$2450000} & \colhead{Velocity (\ms)} & \colhead{Uncertainty (\ms)} & \colhead{Mode}}
\startdata
\label{Xingvels}
4046.34259 & 53.99 & 27.92 & CES-O \\ 
4133.12441 & 53.23 & 21.39 & CES-O \\ 
4398.37834 & 101.38 & 17.20 & CES-O \\ 
4459.23135 & 117.08 & 30.81 & CES-O \\ 
4459.24962 & 32.80 & 27.00 & CES-O \\ 
\hline
5172.26505 & 96.73 & 19.05 & CES-N \\ 
5172.28626 & 76.60 & 15.71 & CES-N \\ 
5172.30759 & 51.67 & 14.37 & CES-N \\ 
5173.27944 & 45.82 & 15.67 & CES-N \\ 
5173.30065 & 56.52 & 16.41 & CES-N \\ 
5173.32198 & 65.67 & 17.79 & CES-N \\ 
5173.34310 & 65.70 & 14.76 & CES-N \\ 
5226.05844 & $-$45.55 & 24.43 & CES-N \\ 
5226.10875 & $-$20.17 & 22.59 & CES-N \\ 
5227.08807 & $-$26.57 & 11.76 & CES-N \\ 
5227.10979 & $-$46.69 & 13.18 & CES-N \\ 
5227.13103 & $-$12.38 & 14.99 & CES-N \\ 
5281.99890 & $-$14.95 & 13.80 & CES-N \\ 
5282.00964 & $-$22.57 & 15.56 & CES-N \\ 
5282.02035 & $-$61.02 & 17.25 & CES-N \\ 
5282.98558 & $-$45.16 & 16.54 & CES-N \\ 
5283.00671 & $-$59.43 & 16.13 & CES-N \\ 
5283.02782 & $-$52.03 & 15.03 & CES-N \\ 
5283.98000 & $-$51.62 & 13.79 & CES-N \\ 
5284.00528 & $-$29.21 & 20.65 & CES-N \\ 
5284.02646 & $-$48.95 & 20.95 & CES-N \\ 
5284.04756 & $-$6.26 & 21.05 & CES-N \\ 
5520.33128 & 39.59 & 17.73 & CES-N \\ 
5520.35299 & 21.56 & 16.20 & CES-N \\ 
5552.23941 & 34.69 & 14.87 & CES-N \\ 
5552.26341 & 24.93 & 21.94 & CES-N \\ 
\hline
5878.34042 & 43.36 & 13.50 & HRS \\ 
5878.36563 & 33.75 & 13.50 & HRS \\ 
5878.38866 & $-$0.73 & 16.05 & HRS \\ 
5878.41177 & 42.81 & 16.18 & HRS \\ 
5879.29416 & 4.27 & 19.96 & HRS \\ 
5905.20117 & 23.16 & 11.56 & HRS \\ 
5905.22422 & 14.68 & 12.04 & HRS \\ 
5905.24720 & 36.90 & 14.27 & HRS \\ 
5905.27024 & 48.98 & 15.64 & HRS \\ 
5907.26067 & 1.38 & 8.18 & HRS \\ 
5907.28402 & $-$8.44 & 9.30 & HRS \\ 
5907.30698 & 8.20 & 10.15 & HRS \\ 
5907.32995 & 7.82 & 9.57 & HRS \\ 
5936.13872 & $-$29.36 & 10.11 & HRS \\ 
5936.16225 & $-$35.52 & 10.34 & HRS \\ 
5936.18572 & $-$32.01 & 10.67 & HRS \\ 
5936.20863 & $-$22.04 & 11.92 & HRS \\ 
5959.11008 & $-$30.57 & 22.97 & HRS \\ 
5959.13303 & $-$43.36 & 20.46 & HRS \\ 
5959.15600 & $-$43.22 & 18.84 & HRS \\ 
5959.17896 & $-$62.76 & 24.07 & HRS \\ 
5960.09450 & $-$49.14 & 13.67 & HRS \\ 
5960.11747 & $-$39.46 & 15.51 & HRS \\ 
5966.04907 & $-$39.47 & 19.32 & HRS \\ 
5966.07211 & $-$31.79 & 20.95 & HRS \\ 
5966.09510 & $-$79.17 & 17.80 & HRS \\ 
5966.11809 & $-$67.76 & 17.67 & HRS \\ 
5997.97169 & $-$66.24 & 10.18 & HRS \\ 
5997.99495 & $-$52.05 & 11.16 & HRS \\ 
5998.01824 & $-$47.66 & 11.05 & HRS \\ 
5998.04152 & $-$60.49 & 10.68 & HRS \\ 
6199.33390 & 51.35 & 6.54 & HRS \\ 
6199.35708 & 45.59 & 7.80 & HRS \\ 
6232.33343 & 54.89 & 5.89 & HRS \\ 
6232.35667 & 61.48 & 6.12 & HRS \\ 
6287.18192 & 30.60 & 9.50 & HRS \\ 
6287.20515 & 35.52 & 10.34 & HRS \\ 
6287.22838 & 32.06 & 9.39 & HRS \\ 
6287.25312 & 47.44 & 11.03 & HRS \\ 
6318.08350 & 31.16 & 11.25 & HRS \\ 
6318.10670 & 13.27 & 11.99 & HRS \\ 
6318.13045 & 21.01 & 12.00 & HRS \\ 
6318.15375 & 25.07 & 11.17 & HRS \\ 
6344.02830 & 11.76 & 8.52 & HRS \\ 
6344.05152 & 23.29 & 9.47 & HRS \\ 
6344.07471 & 22.34 & 9.87 & HRS \\ 
6583.31389 & 22.42 & 10.63 & HRS \\ 
6583.33707 & 35.39 & 11.03 & HRS \\ 
6583.36032 & 20.96 & 11.67 & HRS \\ 
6611.28853 & 18.21 & 7.27 & HRS \\ 
6611.31171 & 20.81 & 8.39 & HRS \\ 
6611.33507 & 7.72 & 8.55 & HRS \\ 
6645.18797 & $-$18.03 & 12.37 & HRS \\ 
6645.21133 & 9.17 & 12.68 & HRS \\ 
6645.23463 & $-$12.05 & 15.41 & HRS \\ 
6645.25792 & $-$22.92 & 16.31 & HRS \\ 
6701.05113 & $-$50.92 & 11.31 & HRS \\ 
6701.07444 & $-$55.62 & 15.95 & HRS \\ 
6728.98753 & $-$39.22 & 16.13 & HRS \\ 
6729.01088 & $-$50.79 & 19.71 & HRS 
\enddata
\end{deluxetable}

\begin{deluxetable}{lcc}
\tabletypesize{\scriptsize}
\tablecolumns{3}
\tablewidth{0pt}
\tablecaption{AAT Radial Velocities for HD 47366}
\tablehead{
\colhead{JD$-$2450000} & \colhead{Velocity (\ms)} & \colhead{Uncertainty(\ms)}}
\startdata
\label{AATvels}
6374.88737  &     $-$4.3  &    1.2  \\
6375.91758  &      1.9  &    1.1  \\
6376.93270  &      0.0  &    1.1  \\
6377.98528  &     $-$2.0  &    1.2  \\
6399.91709  &     13.0  &    1.4  \\
6526.29116  &     70.0  &    1.4  \\
6529.31433  &     85.4  &    1.5  \\
6555.29760  &     63.2  &    1.5  \\
6556.21605  &     67.5  &    1.5  \\
6686.00004  &    $-$17.2  &    1.6  \\
6744.92245  &    $-$42.0  &    1.5  \\
6747.92675  &    $-$46.0  &    1.6  \\
6764.87970  &    $-$40.9  &    1.3
\enddata
\end{deluxetable}


\subsection{AAT Observations}

Since the inner planet has a period near one year, and HD\,47366 is 
quite far south for OAO, we obtained confirmation observations from the 
3.9-metre Anglo-Australian Telescope (AAT).  This multi-telescope 
approach was also critical for the confirmation of the 360-day planet 
HD\,4732b \citep{sato:2013a}.  The UCLES echelle spectrograph 
\citep{diego:90} delivers a resolving power of $R\sim$45,000 with the 
1-arcsecond slit, and has been used for 16 years by the Anglo-Australian 
Planet Search 
\citep[e.g.][]{tinney01,butler01,jones10,142paper,2jupiters}.  
Calibration of the spectrograph point-spread function is achieved using 
an iodine absorption cell temperature-controlled at 
60.0$\pm$0.1$^{\rm{o}}$C.  The iodine cell superimposes a forest of 
narrow absorption lines from 5000 to 6200\,\AA, allowing simultaneous 
calibration of instrumental drifts as well as a precise wavelength 
reference \citep{val:95,butler:1996}.  The result is a precision Doppler 
velocity estimate for each epoch, along with an internal uncertainty 
estimate, which includes the effects of photon-counting uncertainties, 
residual errors in the spectrograph PSF model, and variation in the 
underlying spectrum between the iodine-free template and epoch spectra 
observed through the iodine cell.  HD\,47366 was observed at 13 epochs 
between 2013 March and 2014 April; the AAT velocities are given in 
Table~\ref{AATvels}.

\section{Orbit Fitting and Planetary Parameters}\label{ana}

The collected radial-velocity data for HD 47366 obviously show a variation
with a period near one year. Although the 1yr periodicity and the relatively
low declination of the star allowed us to cover only a half of the cycle, we
can clearly see the periodic variation of the velocity amplitude every year,
indicating the existence of a second period near two years.  Thus we performed
the least-squares orbital fitting by a double Keplerian model.
The orbital parameters and the uncertainties were derived using the Bayesian
Markov Chain Monte Carlo (MCMC) method (e.g., \citealt{ford:2005,gregory:2005,
ford:2007}), following the analysis in \citet{sato:2013b}.
We took account of velocity offsets $\Delta$RV of HIDES-F, CES-O, CES-N, HRS
and AAT data relative to HIDES-S data as free parameters in the orbital fitting.
Extra Gaussian noises $s$ for each of the six data sets were also incorporated as free
parameters, though those for CES-O ($s_3$) and HRS ($s_5$) are fixed to 0
since their RMS scatters to the best-fit orbit are comparable to their measurement
errors.
We generated 5 independent chains having $10^7$ points with acceptance rate of
about 25\%, the first 10\% of which were discarded, and confirmed each
parameter was sufficiently converged based on the Gelman-Rubbin statistic
\citep{gelman:1992}.
We derived the median value of the merged posterior PDF for each parameter
and set 1$\sigma$ uncertainty as the range between 15.87\% and 84.13\% of the PDF.

\begin{figure}
\plotone{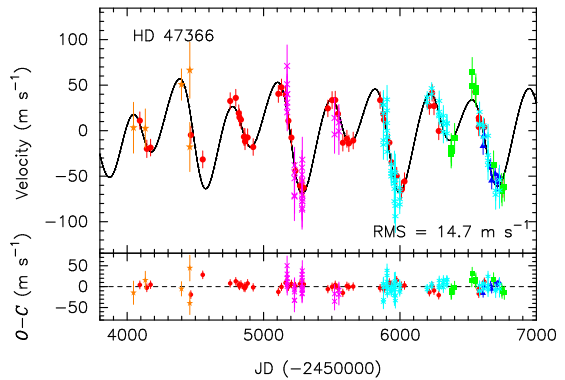}\epsscale{1}
\caption{{\it Upper}: Radial velocities of HD 47366 obtained with HIDES-S (red),
HIDES-F (blue), CES-O (brown), CES-N (magenta), HRS (cyan), and
AAT (green). The error bar for each point includes the extra Gaussian
noise.
The double Keplerian model for the radial velocities is shown by
the solid line. {\it Bottom}: Residuals to the Keplerian fit.}
\label{fig:orbit}
\end{figure}

\begin{figure}\epsscale{1.}
\plotone{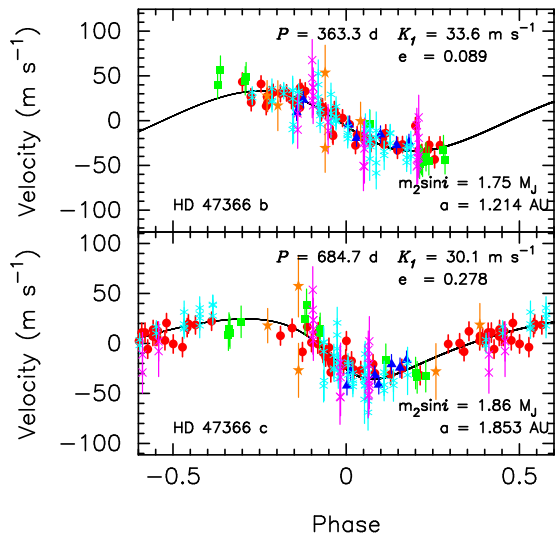}
\caption{Phased radial velocities and the Keplerian models for the inner (upper
panel; signal of the outer planet is removed) and the outer (bottom panel;
signal of the inner planet is removed) planet. The error bar for each point
includes the extra Gaussian noise. The symbols are the same as those in figure \ref{fig:orbit}.}
\label{fig:phase}
\end{figure}

In figure \ref{fig:orbit} and \ref{fig:phase}, we plot the derived Keplerian orbits together with
the radial-velocity points and their measurement errors including the extra Gaussian
noises. The inner planet (planet b) has orbital parameters of period $P_b=363.3^{+2.5}_{-2.4}$ d,
eccentricity $e_b=0.089^{+0.079}_{-0.060}$, minimum mass
$m_b\sin i=1.75^{+0.20}_{-0.17}~\Mjup$, and semimajor axis
$a_b=1.214^{+0.030}_{-0.029}$ AU, and the outer planet (planet c)
has $P_c=684.7^{+5.0}_{-4.9}$ d, $e_c=0.278^{+0.067}_{-0.094}$,
$m_c\sin i=1.86^{+0.16}_{-0.15}~\Mjup$, and $a_c=1.853^{+0.045}_{-0.045}$ AU.
The obtained parameters are listed in Table \ref{tbl:planets}.


\begin{figure}\epsscale{1.2}
\plotone{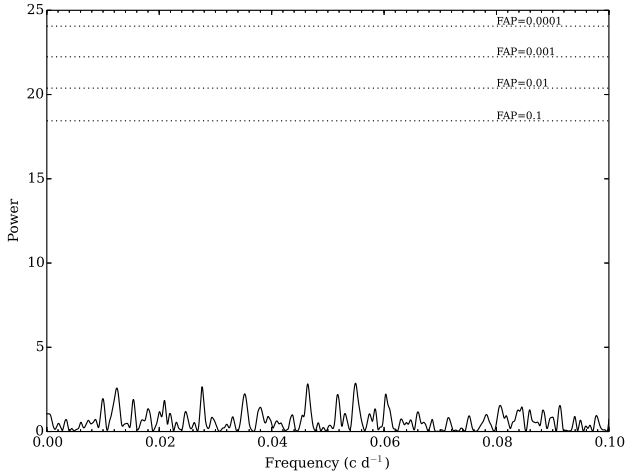}
\caption{Generalized Lomb-Scargle periodogram of the Hipparcos photometric
data of HD 47366. The dotted lines indicate False Alarm Probability ($FAP$)
levels.}
\label{fig:hipp_photo}
\end{figure}

The Hipparcos satellite made a total number of 158 photometric observations for
HD 47366 from 1990 March to 1993 March, and revealed photometric
stability for the star down to $\sigma_{\rm HIP}=0.007$ mag, though the Hipparcos
observations were not contemporaneous with the radial velocity observations.
Figure \ref{fig:hipp_photo} shows the generalized Lomb-Scargle periodogram
\citep{zech:2009} of the Hipparcos photometric data of the star. We did not find any clue
that the radial velocity variations correlates with brightness variations of the star.
Furthermore the rotational period of the star can be estimated to be shorter than about
86 days based on its radius $R=7.3~\Rsun$ and its projected rotational
velocity $v\sin i=4.3~\kms$ (see Table \ref{tbl:star}).  This is much shorter than
either the 363-day or the 685-day period observed in radial velocity. Thus rotational
modulation of spots on the stellar surface is not considered to be a viable explanation
for the observed radial-velocity variations.

We also performed spectral line shape analysis for the star following the method
in \cite{sato:2007}.  Cross-correlation profiles between pairs of stellar templates,
each of which were extracted from five I$_2$-superposed spectra at phases of velocity
$\sim30~\ms$ (phase 1), $\sim-10~\ms$ (phase 2), and $\sim-50~\ms$ (phase 3)
by using the method of \cite{sato:2002}, were derived for about 110 spectral
segments (4--5${\rm \AA}$ width each). Three bisector quantities of the cross-correlation
profiles, BVS, BVC, and BVD, were then calculated, which are the velocity difference
between the two flux levels of the bisector, the difference of the BVS of the upper and
lower halves of the bisector, and the average of the bisector at three flux levels,
respectively. The flux levels of 25\%, 50\%, and 75\% of each cross-correlation
profile were used to calculate the above three bisector quantities. As a result, we
obtained BVS$= 6.6\pm5.5~\ms$, BVC$=-2.5\pm3.0~\ms$, and
BVD$= -86.1\pm10.8~\ms$ for the cross-correlation profile between phase 1 and 3,
and BVS$= -4.1\pm4.9~\ms$, BVC$=3.1\pm2.6~\ms$, and BVD$= -35.1\pm11.6~\ms$
for the one between phase 2 and 3.
The BVD values are consistent with the velocity differences between the phases,
and the BVS and BVC values are much smaller than the BVD values.
Therefore we conclude that the observed radial velocity variations are not
originated from distortion of the spectral lines but from parallel shifts of them
as expected for orbital motion.

\section{Dynamical Stability}\label{dynamics}

The orbital parameters obtained in Section \ref{ana} were derived by fitting a double
Keplerian model to the radial velocity data, which does not guarantee that the orbits
are stable over long periods of time. In order to investigate the orbital stability of the system
and further constrain orbital parameters, we performed dynamical analysis for the
system. We used a fourth-order Hermite scheme for the numerical integration
\citep{kokubo:1998}.

\begin{figure}\epsscale{1.2}
\plotone{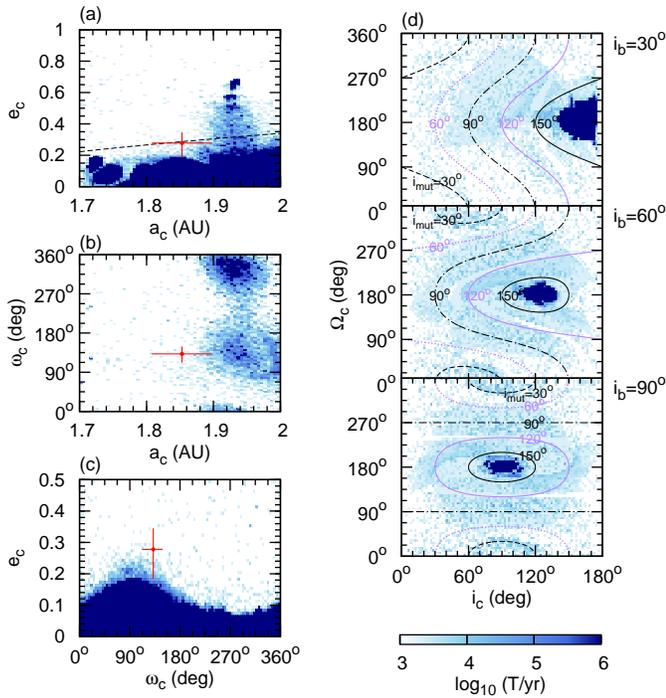}
\caption{Life-time of the system. Time elapsing before the semimajor axis of one planet deviates by 10\% from
its initial value is shown. Since the extent of our numerical integrations is $10^6$yr, the life time of $10^6$yr
here means the system is stable over $10^6$yr (dark-blue regions).
Red cross represents the best-fitted $a_c$ and $e_c$ to the radial-velocity data
with their 1$\sigma$ errors.
(a) The diagram of $e_c$ versus $a_c$. (b) The diagram of argument of pericenter $\omega_c$ versus semimajor axis.
(c) The $e_c$-$\omega_c$ diagram. (d) The diagram of longitude of ascending node $\Omega_c$ versus
inclination from the line of sight $i_c$, together with the contours of the mutual inclination. 
From top to bottom, the cases for $i_b=30^{\circ}$,  $60^{\circ}$, and $90^{\circ}$ are shown. 
The absolute masses of planet b and c are inversely proportional to $\sin i_b$ and $\sin i_c$, respectively.
\label{fig:lifetime}}
\end{figure}

Figure \ref{fig:lifetime} shows the life-time of the system calculated by the $10^6$ yr
integrations. The life-time here is defined as the time elapsing before the
semimajor axis of one planet deviates by 10\% from its initial value.
All of the initial orbital parameters for the orbital integrations are fixed to the best-fit
ones derived in Section \ref{ana} except for those shown in each axis of each panel. 
Figures \ref{fig:lifetime}({\it a}), ({\it b}), and ({\it c}) shows the life-time of the system
in the ($a_c, e_c$), ($a_c, \omega_c$), and ($\omega_c, e_c$) plane, respectively.
In the panels, prograde and edge-on orbits ($i_b=i_c=90^{\circ}$) are assumed. 
The dashed-line in panel ({\it a}) shows the orbit-crossing boundary where the apocenter
distance of planet b is equal to the pericenter distance of planet c. The planet b and
c actually have almost crossing orbits with each other in the case of the best-fit orbits
(marked by red cross in each panel), and as shown in the figure, they are unstable.
Orbits above the boundary line easily become unstable within 1000 yr. 
Stable orbits appearing above the boundary are in the 2:1 mean-motion resonance.
If $a_c$ is near 1.93~AU with an appropriate argument of pericenter, the orbits are stable
with up to $e_c\sim0.7$ over $10^6$ yr thanks to the resonance, though this $a_c$ value
is beyond the 3$\sigma$-range of $a_b/a_c (=0.6555^{+0.0041}_{-0.0043}$) that is well
determined by our radial-velocity data (panel (b)). Or, if $e_c$ is smaller than $\sim$0.15
($1.4\sigma$ away from the best-fit value),
the orbits are also stable even with the best-fit $a_c$ (1.85AU) in prograde
configuration (panel (c)).

If the mutual inclination $i_{\rm mut}$ between the two planets is higher than
$160^{\circ}$, the orbit becomes stable in wider parameter range. 
Figure~\ref{fig:lifetime}~({\it d}) shows the stability map in the plane of inclination ($i_c$)
and ascending node  ($\Omega_c$) of planet c (ascending node of planet b is
set to $\Omega_b=0$). 
The mutual inclination of planets depends on $i_b$, $i_c$ and $\Omega_c-\Omega_b$ as
$\cos i_{\rm mut}=\cos i_b\cos i_c +\sin i_b \sin i_c \cos  (\Omega_c-\Omega_b)$,
and its contours are shown by black or purple lines in the figure. 
Since the absolute masses of the planets depend on the angle from the line of sight, the cases of
three different inclinations for planet b, $i_b=30^{\circ}$, $60^{\circ}$, and $90^{\circ}$, are shown.
The absolute mass of planet c is also changed according to $i_c$. 
The absolute masses of planet b and c are inversely proportional to $\sin i_b$ and $\sin i_c$, respectively.
As seen in the figure, the orbits with $i_{\rm mut} \sim 180^{\circ}$ (i.e., mutually retrograde) are
stable in all cases.
This is because planet encounters occur in a shorter time with higher mutual velocities in the
retrograde system compared to the prograde system. 
This is similar to the case of the two brown dwarf candidates orbiting BD+20 2457, in which the system is
unstable for the prograde orbits, while it is stable for the retrograde orbits \citep{horner:2014}.

\begin{figure}\epsscale{1.2}
\plotone{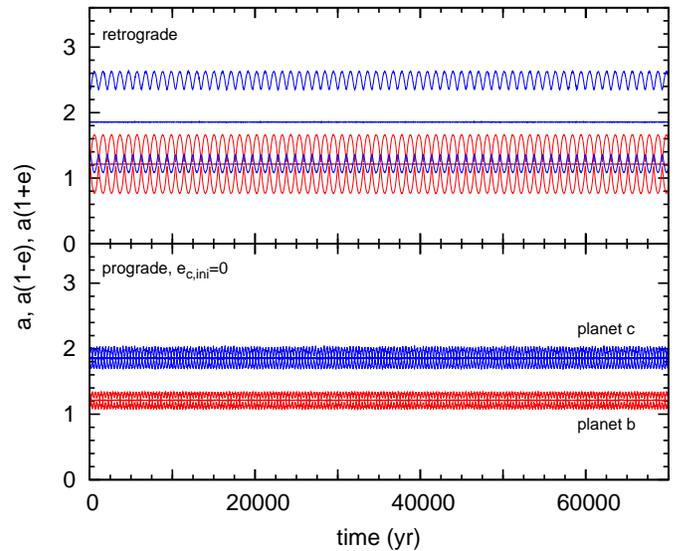}
\caption{Evolution of semimajor axis, pericenter distance, and apocenter distance of planet b (red lines)
and c (blue lines) assuming $i_b=i_c=90^{\circ}$. {\it Upper}: retrograde coplanar case with best-fitting orbital elements.
{\it Bottom}: prograde coplanar case, but $e_c=0$ is assumed. 
\label{fig:orbitcalc}}
\end{figure}

Figure \ref{fig:orbitcalc} shows the $7 \times 10^4$ yr evolution of the system. 
The upper panel shows the evolution of the coplanar retrograde system with the best-fit orbital
parameters derived in section \ref{ana}. Although the outer planet violates the orbit of the inner
planet, they are stable over $10^6$ yr. 
The bottom panel shows a prograde case with the same orbital parameters except for assuming
$e_{\rm c, ini}=0$. As seen in the figure, if the eccentricity of the outer planet is small, the system can
keep stable orbits, while the system with the outer planet in eccentric orbit quickly
causes orbital crossing and becomes unstable (see figure \ref{fig:lifetime}).

\begin{figure}\epsscale{1.2}
\plotone{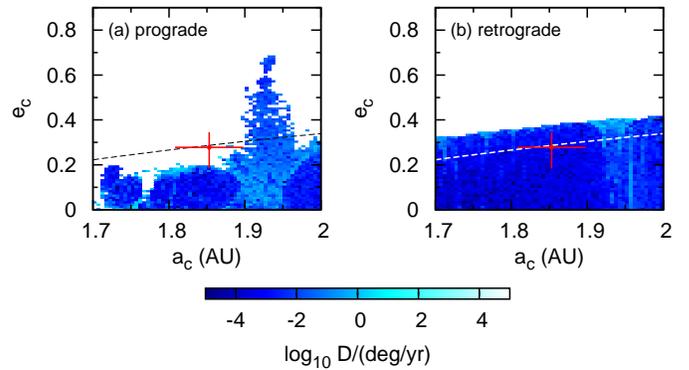}
\caption{Mean motion diffusion $D$ (stability index) of the HD47366 system. 
The best-fit orbital parameters are used as the initial parameters for the orbital integrations except for semimajor
axis and eccentricity of planet c, whose best-fitted values are marked by red crosses
with their 1$\sigma$ errors. 
The absolute mass of planets used in the simulations are $m_b=1.75~\Mjup$ and $m_c=1.86~\Mjup$
($i_b=i_c=90^{\circ}$).
The dashed lines show the orbit-crossing boundary where  the apocenter
distance of planet b is equal to the pericenter distance of planet c. 
(a) prograde coplanar configuration. (b) retrograde coplanar configuration. 
\label{fig:freqmap}}
\end{figure}

Figure \ref{fig:freqmap} shows maps of the stability index $D=|\langle n_2 \rangle -\langle n_1 \rangle |$
for the system. 
From  5000 yr numerical integration, we obtain an average of mean motion $\langle n_1 \rangle $
of planet c and subtract it from an average of mean motion $\langle n_2 \rangle $ obtained 
in the next 5000 yr orbit (see \cite{couetdic:2010} for the details of the stability analysis).
The left and right panels are the maps for prograde and retrograde configuration, respectively.
From comparison with panel ({\it a}) of Figure \ref{fig:lifetime}, we find that the system is regular
when $\log_{10} D \le -3$.
In the case of the retrograde orbit, the orbit is stable even if the eccentricity of the planet c exceeds
the orbit-crossing boundary (dashed line) by $\sim$0.1.

\section{Discussion and Summary}\label{summary}

We have reported the detection of a double giant-planet system around the K1 giant
HD 47366 ($M_{\star}=1.81\pm0.13~M_{\odot}$, $R_{\star}=7.30\pm0.33~R_{\odot}$)
from precise radial-velocity measurements at OAO, Xinglong, and AAT.
The inner (planet b) and outer planet (planet c) have minimum mass of
$1.75^{+0.20}_{-0.17}~\Mjup$ and
$1.86^{+0.16}_{-0.15}~\Mjup$, semimajor axis of $1.214^{+0.030}_{-0.029}$ AU
and $1.853^{+0.045}_{-0.045}$ AU, and eccentricity of $0.089^{+0.079}_{-0.060}$
and $0.278^{+0.067}_{-0.094}$, respectively, which were derived by fitting a double
Keplerian model to the radial-velocity data. The period ratio of the two planets
is slightly smaller than 2, and thus the system adds to the growing population
of such multi-giant-planet systems with small orbital separation around evolved
intermediate-mass stars.

Dynamical analysis for the system revealed, however, that the best-fit orbits in
prograde configuration are unstable.
We found that it is stable in the following cases;
1) the two planets are in the 2:1 mean-motion resonance;
2) the eccentricity of planet c is less than $\sim$0.15;
3) mutual inclination of two the planets is larger than $160^{\circ}$ (i.e. retrograde configuration).

If the two planets are in the stable 2:1 resonance, the ratio of the semimajor axes
should be $a_b/a_c\sim0.63~(a_c\sim1.93{\rm AU})$ and the eccentricity of the planet c
should be $0.5\lesssim e_c \lesssim 0.7$ or $0\le e_c \lesssim 0.2$ (see section \ref{dynamics}
and darkest-blue regions in the panel (a) of figure \ref{fig:lifetime}).
On the other hand, $a_b/a_c$ is well determined from our observations to be
$a_b/a_c=0.6555^{+0.0041}_{-0.0043}$ and $e_c$ is determined to be
$e_c=0.278^{+0.067}_{-0.094}$. Both of these values deviate from those
for the high-eccentricity case with more than 3$\sigma$, and thus
from the view point of the observational data, it is less likely that the planets are in
the high-eccentricity resonance. We can not completely reject the possibility of the low-eccentricity
case. However, considering the semimajor-axis ratio, it is also less likely that
the planets are in the resonance.

It is more plausible that the eccentricity of the planet c is less than $\sim$0.15. The value is
just $1.4~\sigma$ from the best-fit value, and the best-fit orbit is consistent with a
circular one to the 3$\sigma$ level. Thus, it is reasonable to think that the two planets actually
have nearly circular orbits and thus they are stable.
In this case, the period ratio of the planets is slightly smaller than 2, and they are not in
2:1 resonance.

The third possibility is the most extreme one. Basically we cannot know inclination or
ascending node of a planetary orbit by radial velocity measurements, and thus we cannot
discriminate between prograde and retrograde orbits observationally.
\cite{gayon:2008} and \cite{gayon:2009} studied mutually retrograde configurations for
some multi-planet systems around solar-type stars and showed that the
configurations are dynamically stable and consistent with the radial-velocity observations.
Recently, \cite{horner:2014} studied the dynamical stability of a double brown-dwarf
system around the giant BD+20 2457 \citep{niedzielski:2009}, and showed that the
prograde orbits best-fitted to the radial-velocity data are unstable,
while those in retrograde configuration are stable.
The HD 47366 system is similar to the case of the BD+20 2457 system. Our
dynamical analysis showed that the coplanar retrograde configuration for
the HD 47366 system is stable in a wider range of orbital parameters including the eccentric
ones which are unstable for prograde configuration (see figure \ref{fig:orbitcalc}).
Confirmation of the high eccentricity of the outer planet by continuous observations will
strongly support the retrograde hypothesis for the HD 47366 system, though the formation
mechanism of such a mutually-retrograde system is largely unknown. 
In any of the above cases, the planetary orbits can be altered by close encounter
of the planets in relatively short period of time, and thus more detailed dynamical
modeling for the observed radial-velocity variations is necessary in order to provide
a definitive orbital solution.

The stellar and planetary parameters of HD 47366 system are similar to those of
$\eta$ Cet system; two massive planets ($m_b \sin i=2.6~\Mjup$,
$m_c \sin i=3.3~\Mjup$), with periods of $P_b = 407$ days and
$P_c = 740$ days and eccentricities of $e_b = 0.12$ and $e_c = 0.08$
orbit around a K giant star with a mass of 1.7 $M_{\odot}$ and a radius of 14.3$R_{\odot}$
\citep{trifonov:2014}. \cite{trifonov:2014} revealed that, for a coplanar configuration, the system
can be stable by being trapped
in an anti-aligned 2:1 mean motion resonance in a region with moderate $e_b$ eccentricity,
which lies about 1$\sigma$ away from the best-fit Keplerian orbits. A much larger non-resonant
stable region also exists in low-eccentricity parameter space, though it appears to be much
farther from the best fit than the 2:1 resonant region.
This is in contrast to the case of HD 47366 that low eccentricities of the planets are more
plausible than 2:1 resonant configuration. Another possible 2:1 resonant system around
an evolved star is 24 Sex, in which two massive planets ($m_b \sin i=1.99~\Mjup$,
$m_c \sin i=0.86~\Mjup$), with periods of $P_b = 452.8$ days and $P_c = 883.0$
days and eccentricities of $e_b = 0.09$ and $e_c = 0.29$ orbit around a G-type subgiant
with a mass of 1.54$M_{\odot}$ and a radius of 4.9$R_{\odot}$ \citep{johnson:2011}.
The period ratio of the planets is the closest to 2 among the three systems.
\cite{wittenmyer12} revealed that the best-fit orbit of the system is mutually crossing and it is
only dynamically feasible if the planets are in 2:1 resonance that can protect them
from close encounters. As described, all the above three systems are near the 2:1 resonance,
but they show different properties in their orbital configuration. Further analysis and comparison
of the systems would help us understand dynamics of planetary systems in more detail.

It is unknown why multi-giant-planet systems with small orbital separation are preferentially
found around evolved intermediate-mass stars (see figure \ref{period-ratio}). It may be a
primordial property of planets around intermediate-mass stars that could be an outcome
of planet formation or an acquired one as a result of orbital evolution caused by
stellar evolution (stellar tide and mass loss) of central stars. In the case of HD 47366, if the
star is a helium-core burning star that has passed through the RGB tip (see section \ref{stpara}),
the planetary orbits could have been migrated outward by $\sim10\%$ ($\sim$0.1AU) during
the RGB phase because of the effect of mass loss of the central star \citep{kunitomo:2011, villaver:2014}.
However, the effect works for both of the two planets and thus it could not result in making
the orbital separation smaller. The star might have harbored a third planet that was engulfed
and possibly partially responsible for the current orbital configuration, though the star does
not exhibit a possible signature of the planet engulfment such as an overabundance of lithium
\citep[e.g.][]{siess:1999, adamow:2012}; \cite{liu:2014} obtained the lithium abundance of
$A({\rm Li})=0.38$\footnote{$A({\rm Li})=\log n_{\rm Li}/n_{\rm H}+12$} for HD 47366 suggesting
that the star is an Li-depleted giant.
Considering also that many of the systems with small orbital
separation are found around less evolved subgiants, the effect of stellar evolution may
be less significant compared to that of stellar mass.
Unfortunately the ongoing Doppler planet searches cannot clearly
discriminate between the above two factors, stellar mass and stellar evolution, because the
targets of the searches are at once evolved and intermediate-mass stars.
Investigating planets around intermediate-mass main-sequence stars or those around
evolved low-mass ($\sim 1~M_{\odot}$) stars would help clarify each effect separately.


\begin{deluxetable*}{lrr}
\tablecaption{Orbital Parameters for HD\,47366\label{tbl:planets}}
\tablewidth{0pt}
\tablehead{
\colhead{Parameter} & \colhead{HD\,47366 b} & \colhead{HD\,47366 c}}
\startdata
  \label{tbl-planets}
Period $P$ (days)                            &  363.3$^{+2.5}_{-2.4}$        & 684.7$^{+5.0}_{-4.9}$     \\
RV semiamplitude $K_1$ (\ms)                           &  33.6$^{+3.6}_{-2.8}$      & 30.1$^{+2.1}_{-2.0}$    \\
Eccentricity $e$                                   &  0.089$^{+0.079}_{-0.060}$ & 0.278$^{+0.067}_{-0.094}$\\
Longitude of periastron $\omega$ (deg)                        &  100$^{+100}_{-71}$     & 132$^{+17}_{-20}$    \\
Periastron passage $T_p$    (JD$-$2450000)             & 122$^{+71}_{-55}$         & 445$^{+55}_{-62}$       \\
Minimum mass $m_2\sin i$ ($\Mjup$)                    & 1.75$^{+0.20}_{-0.17}$                & 1.86$^{+0.16}_{-0.15}$              \\
Semimajor axis $a$ (AU)                                           & 1.214$^{+0.030}_{-0.029}$     & 1.853$^{+0.045}_{-0.045}$      \\
\multicolumn{1}{l}{Semimajor axis ratio $a_{\rm b}/a_{\rm c}$} & \multicolumn{2}{c}{0.6555$^{+0.0041}_{-0.0043}$} \\
\multicolumn{1}{l}{Extra gaussian noises for HIDES-S $s_1$ (\ms)} & \multicolumn{2}{c}{8.4$^{+1.4}_{-1.2}$} \\
\multicolumn{1}{l}{Extra gaussian noises for HIDES-F $s_2$ (\ms)} & \multicolumn{2}{c}{8.5$^{+4.7}_{-3.1}$} \\
\multicolumn{1}{l}{Extra gaussian noises for CES-O $s_3$ (\ms)} & \multicolumn{2}{c}{0 (fixed)}\\
\multicolumn{1}{l}{Extra gaussian noises for CES-N $s_4$ (\ms)} & \multicolumn{2}{c}{13.1$^{+5.8}_{-7.1}$} \\
\multicolumn{1}{l}{Extra gaussian noises for HRS $s_5$ (\ms)} & \multicolumn{2}{c}{0 (fixed)} \\
\multicolumn{1}{l}{Extra gaussian noises for AAT $s_6$ (\ms)} & \multicolumn{2}{c}{15.8$^{+5.6}_{-4.5}$} \\
\multicolumn{1}{l}{Velocity offset of HIDES-F $\Delta$RV$_{2-1}$ (\ms)} & \multicolumn{2}{c}{38.3$^{+5.2}_{-5.2}$} \\
\multicolumn{1}{l}{Velocity offset of CES-O $\Delta$RV$_{3-1}$ (\ms)} & \multicolumn{2}{c}{50$^{+11}_{-11}$} \\
\multicolumn{1}{l}{Velocity offset of CES-N $\Delta$RV$_{4-1}$ (\ms)} & \multicolumn{2}{c}{25.3$^{+4.7}_{-4.7}$} \\
\multicolumn{1}{l}{Velocity offset of HRS $\Delta$RV$_{5-1}$ (\ms)} & \multicolumn{2}{c}{14.5$^{+2.4}_{-2.5}$} \\
\multicolumn{1}{l}{Velocity offset of AAT $\Delta$RV$_{6-1}$ (\ms)} & \multicolumn{2}{c}{20.5$^{+5.6}_{-5.3}$} \\
\multicolumn{1}{l}{Number of data of HIDES-S $N_1$} & \multicolumn{2}{c}{50} \\
\multicolumn{1}{l}{Number of data of HIDES-F $N_2$} & \multicolumn{2}{c}{7} \\
\multicolumn{1}{l}{Number of data of CES-O $N_3$} & \multicolumn{2}{c}{5} \\
\multicolumn{1}{l}{Number of data of CES-N $N_4$} & \multicolumn{2}{c}{26} \\
\multicolumn{1}{l}{Number of data of HRS $N_5$} & \multicolumn{2}{c}{60} \\
\multicolumn{1}{l}{Number of data of AAT $N_6$} & \multicolumn{2}{c}{13} \\
\multicolumn{1}{l}{RMS (\ms)} & \multicolumn{2}{c}{14.7}
\enddata
\tablecomments{Velocity offsets are the values relative to HIDES-S data.}
\end{deluxetable*}

%


\acknowledgments

This research is based on data collected at Okayama Astrophysical
Observatory (OAO), which is operated by National Astronomical
Observatory of Japan, Xinglong station, which is operated by
National Astronomical Observatory of China, and at Australian
Astronomical Observatory.
We are grateful to all the staff members of the observatories for their
support during the observations.
This work was partially Supported by the Open Project Program of the
Key Laboratory of Optical Astronomy, National Astronomical Observatories,
Chinese Academy of Sciences.
We thank students of Tokyo Institute
of Technology and Kobe University for their kind help for the observations
at OAO.
BS was partly supported by MEXT's program "Promotion of Environmental
Improvement for Independence of Young Researchers" under the Special
Coordination Funds for Promoting Science and Technology,  and by
Grant-in-Aid for Young Scientists (B) 17740106 and 20740101 and
Grant-in-Aid for Scientific Research (C) 23540263 from the
Japan Society for the Promotion of Science (JSPS).
MN is supported by Grant-in-Aid for Young Scientists (B)
21740324 and HI is supported by Grant-In-Aid for Scientific Research
(A) 23244038 from JSPS.
This work was partly funded by the National Natural Science Foundation
of China under grants 111173031, 1233004, 11390371, as well as the
Strategic Priority Research Program "The Emergence of Cosmological
Structures" of the Chinese Academy of Sciences, Grant No. XDB09000000,
and by the JSPS under Grant-in-Aid for Scientific Research (B)
17340056 (H.A.) and grant 08032011-000184 in the framework of the Joint
Research Project between China and Japan.
This research has made use of the SIMBAD database, operated at
CDS, Strasbourg, France.


\end{document}